\documentclass[final,3p,times,twocolumn]{elsarticle}

\usepackage{graphicx}
\usepackage{epsfig}

\usepackage{amssymb}
\usepackage{lineno}
\usepackage[usenames]{color}
\usepackage[normalem]{ulem}

\input colordvi

\definecolor{orange}{rgb}{1,0.5,0}

\journal{Elsevier}

\begin{document}

\begin{frontmatter}

\title{Angle-resolved X-ray photoemission electron microscopy}

\author[elettra]{T.~O.~Mente\c{s}}
\ead{mentest@elettra.trieste.it}

\author[elettra]{A.~Locatelli}
\ead{locatellia@elettra.trieste.it}

\address[elettra]{Sincrotrone Trieste S.C.p.A., Basovizza-Trieste 34149, Italy}

\begin{abstract}
Synchrotron based photoemission electron microscopy with energy filter combines real space imaging with microprobe diffraction ($\mu$-ARPES), giving access to the local electronic structure of laterally inhomogeneous materials. We present here an overview of the capabilities of this technique, illustrating selected applications of angle resolved photoemission electron microscopy and related microprobe methods. In addition, we report the demonstration of a darkfield XPEEM (df-XPEEM) imaging method for real space mapping of the electronic structure away from $\Gamma$ at a lateral resolution of few tens of nm. The application of df-XPEEM to the (1$\times$12)-O/W(110) model oxide structure shows the high sensitivity of this technique to the local electronic structure, allowing to image domains with inequivalent adsorption site symmetry. Perspectives of angle resolved PEEM are discussed.
\end{abstract}


\end{frontmatter}



\section{Introduction}

X-ray photoemission electron microscopy with energy filter, or spectroscopic PEEM, is a synchrotron-based technique for the spectro-microscopic characterization of laterally inhomogeneous surfaces and interfaces~\cite{BauRPP57}. By exploiting the high brightness of third generation sources, spectroscopic PEEM probes the local chemistry and electronic structure of materials at the mesoscopic scale~\cite{SchmidtSRL98, locaJPCM08}. This technique has found application in the investigation of various phenomena from surface chemical reactions~\cite{LocaCHEM06} to thin film growth on oxide surfaces~\cite{MenPRB76, LocaJPCM07}, from quantum dots \cite{Bia11,RatSML06} and nanotubes~\cite{SuzAPL04} to chemically modified semiconductor surfaces~\cite{SmdPRL07}. 

Prominent examples addressing from a microscopy standpoint the study of electronic structure properties are relatively few in an otherwise very active field. One such study has employed core level and valence band photoemission imaging to monitor the oxidation of ultra-thin Mg and Al films and its relationship with the quantum-well states resulting from electron confinement~\cite{aballe04,aballe10}. By quantifying the correlation between reactivity and the thickness dependent density of states at the Fermi level, such experiments led to the understanding of the oxidation mechanisms based on electron density decay length effects~\cite{binggeli06}. A similar study using valence band XPEEM investigated the electron and photon beam assisted oxidation of the Ag(111) surface~\cite{GntAPL08, GntCPC10}. Other experiments focused on circular dichroism in valence band photoemission from an Ag monolayer grown on Ru(0001), a phenomenon which has been explained by a lowering of the system symmetry due to the incidence direction of the photon beam~\cite{mascaraque11}. Favored by recent instrumentation developments, the field of application of valence-band PEEM may significantly expand by the use of laboratory based conventional UV sources~\cite{fujiPRB09}.

\begin{figure*}
\begin{center}
\includegraphics[width=15cm]{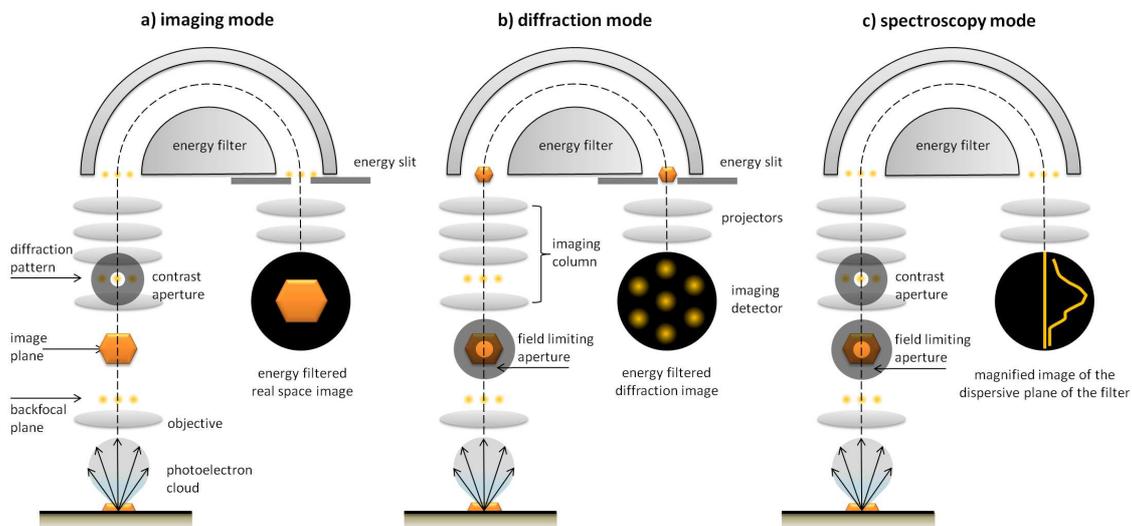}
\caption{Electron optics configuration of the SPELEEM microscope in three different operation modes: a) spectroscopic imaging; b) spectroscopic diffraction imaging; c) micro spectroscopy operation.}
\label{fig:operationModes}
\end{center}
\end{figure*}
 
Along with spectroscopy, most XPEEM microscopes in use nowadays allow to perform microprobe angle-resolved photoemission ($\mu$-ARPES) and photoelectron diffraction (PhD) measurements. The unprecedented scientific interest recently sparked by graphene stimulated numerous such $\mu$-ARPES studies, either focusing on graphene epitaxially grown on transition metal substrates, such as Pt~\cite{SuttPRB80} and Ru~\cite{suttNL09}, or on exfoliated graphene on SiO$_2$ \cite{knox08,knox11}. 
Due to the modest energy resolution (0.3 eV), the application of the microprobe technique has been confined to a small number of laterally inhomogeneous systems, while high energy resolution ARPES facilities are preferred whenever uniform films can be obtained~\cite{ViroPRB08}. 
Below, we will demonstrate a novel approach to valence band PEEM using dark-field methods, opening up the possibility to carry out off-normal, angle resolved measurements at a lateral resolution nearing that of common XPEEM operation. 
 
The dark-field imaging is a widely used concept in a variety of microscopy techniques~\cite{ottensmeyer69,pashley65,graef01,kaulich11}. Its underlying principle is the selection and enhancement of specific scattered or diffracted beams for image formation, while the main probing beam is blanked out. In darkfield low-energy electron microscopy (df-LEEM), an aperture carefully-positioned in the diffraction plane selects a specific secondary diffraction beam for imaging the lateral distribution of the corresponding surface phase~\cite{BauRPP57, altman10}. Applications of  df-LEEM allow to distinguish symmetry properties, e.g.\@ for determination of surface termination and stacking faults~\cite{gabaly07}, or adsorption-site symmetry~\cite{stojicPd}. Recently, darkfield LEEM based on electron exchange scattering from magnetic lattices was successfully used to image antiferromagnetic domains in NiO~\cite{menon11}. 

An analogous approach to df-LEEM can be in principle carried out in the case of XPEEM. The simultaneous use of an energy filter and contrast aperture is necessary to select a well-defined photoemission angle and kinetic energy, which leads to real space XPEEM imaging of the corresponding features in the electronic structure. 
In this work we present an overview of the current capabilities of electronic structure mapping and imaging using the XPEEM. After a short review of recent applications of $\mu$-ARPES on exfoliated graphene, we will introduce the methodology of darkfield operation in angle resolved XPEEM imaging, reporting a first demonstration of the capabilities the method and illustrating our results on (1$\times$12)-O/W(110) in connection with complementary df-LEEM data. We will conclude by discussing future applications and perspectives.

\section{Experimental Setup}

The measurements were performed with the spectroscopic photoemission and low-energy electron microscope (SPELEEM) installed at the Nanospectroscopy beamline of the Elettra synchrotron laboratory (Italy)~\cite{locatelli06}. In the SPELEEM, the specimen can be probed either with low energy electrons (0-750 eV) provided by an LaB$_6$ source, or using  monochromatized soft X-rays in the range 40-1000 eV. The combination of LEEM with energy filtered XPEEM enables to carry out a variety of complementary analytical surface characterization methods with both chemical and structural sensitivity~\cite{SchmidtSRL98}.

The SPELEEM is equipped with a hemispherical bandpass energy filter, which is normally operated at a pass energy of 908 eV. Spectroscopic operation can be implemented either in real space (spectroscopic imaging, or XPEEM) or diffraction imaging mode (microprobe LEED, microprobe angle-resolved photoemission, microprobe photoelectron diffraction), or spectroscopy mode (microprobe-XPS). Schematic diagrams representing the electron optical configuration of the microscope are shown in Fig.~\ref{fig:operationModes}. As can be seen, both in the microprobe diffraction and spectroscopy operation modes, the use of a field limiting aperture restricts the measurements to micron sized surface areas. As reported earlier, the lateral  resolution of the SPELEEM microscope is about 10~nm in LEEM and 30~nm in XPEEM mode; the best energy resolution, 0.2~eV in the microspectroscopy mode, increases to 0.3~eV in both the imaging and diffraction operation~\cite{locatelli06, mentes11}.

\begin{figure}
\begin{center}
\includegraphics[width=7.5cm]{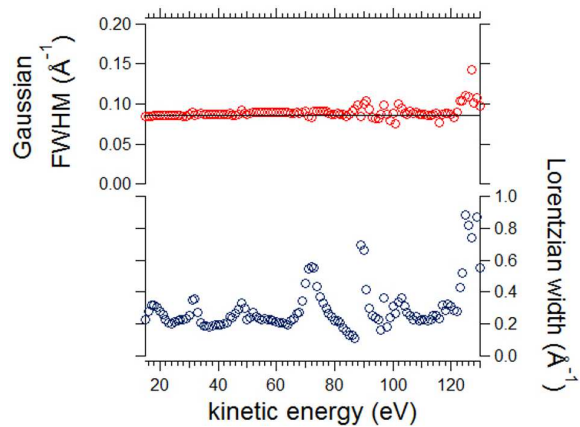}
\caption{The full-width half maximum of the (00) low-energy electron diffraction spot from a defect-free area on a HOPG surface. The error bars represent the uncertainty in the Gaussian fit to the spot profile. The data points are averages FWHM values along perpendicular directions, which show slight differences due to a residual astigmatism in the diffraction plane.}
\label{fig:transferWidth}
\end{center}
\end{figure}

When operated as a LEED instrument, the SPELEEM has a transfer width of about 12 nm. This value was estimated by measuring the FWHM of the Gaussian component of the primary diffraction beam on a defect free highly-oriented pyrolytic graphite (HOPG) specimen. Instrumental and detector broadening, assumed to be Gaussian, were separated from other effects such as inelastic broadening, broadening due to defects, roughness etc.\@, which were taken into account by a Lorentzian curve in the the peak fit. Note that the Lorentzian has a much weaker amplitude compared to the Gaussian. As can be seen from Fig.~\ref{fig:transferWidth}, the Gaussian peak width is nearly constant within the kinetic energy range 15-130~eV. On the other hand, the Lorentzian width instead shows significant changes, reaching maxima away from the Bragg peaks, which is consistent with studies on diffraction profiles from rough surfaces~\cite{LocaACSN10}. The data were acquired for an emission current of  0.1 $\mu A$ of the electron source.

Due to the difficulty of finding extremely sharp features in the angle resolved spectra 
of most materials, we could not follow the same procedure to determine the angular resolution of the microscope in the ARPES operation mode. Nonetheless, the transfer width obtained from the measurement shown in Fig.~\ref{fig:transferWidth} provides a worst case estimate of the angular resolution of the microscope. In fact, the transfer width is largely 
affected by the quality of the electron source, and was found to decrease 
notably with increasing emission current. Decreasing the filter pass energy might be used to improve the energy resolution, but has the drawback of deteriorating the angular resolution, so that, in practical applications, a compromise has to be found case by case.

An important parameter in the darkfield operation is the angular acceptance of the diffraction-plane aperture. The SPELEEM setup currently has three circular pinhole apertures installed at the diffraction plane. The diameters of the apertures in reciprocal space units are 1.69 , 0.51 and 0.34 \AA$^{-1}$, for the large, medium and small apertures respectively. These numbers essentially define the k-space resolution in the darkfield mode, and are significantly larger than the angular resolution of the microscope displayed in Fig.~\ref{fig:transferWidth}.

\section{Microprobe ARPES}

\begin{figure*}
\begin{center}
\includegraphics{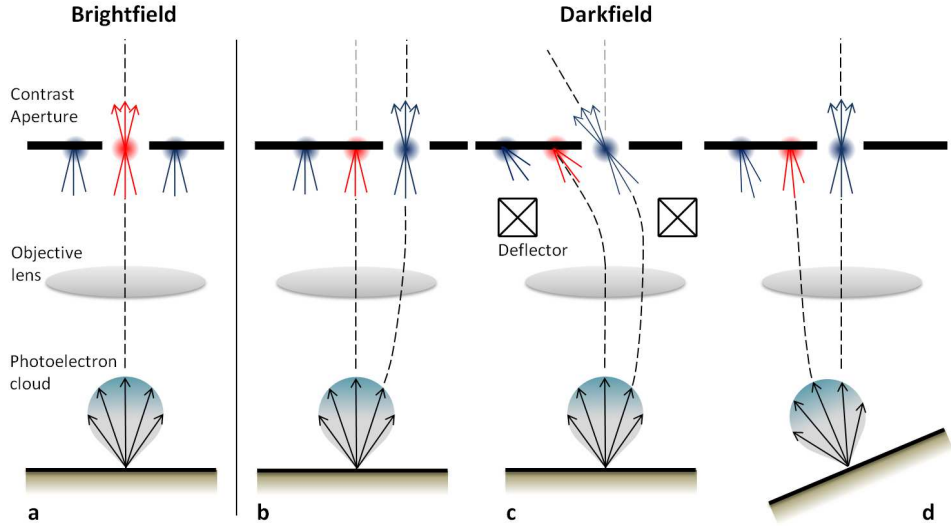}
\caption{a) Brightfield PEEM operation with the contrast aperture centered along the instrument optical axis; b) df-XPEEM operation by displacing the contrast aperture; c) Deflecting the photoelectron beam after the objective lens d) tilting the sample; for explanation see text.}
\label{fig:dfSketch}
\end{center}
\end{figure*}

\subsection{Methodology of diffraction imaging}

In the $\mu$-ARPES mode, the diffraction pattern, formed at back focal plane of the objective lens, is imaged (see Fig.~\ref{fig:operationModes}). The probed area is selected by positioning an aperture in one of image planes along the optical column of the instrument. This ensures that only photoelectrons emitted from the area of interest contribute to the formation of the diffraction pattern. Spectroscopy measurements are carried out after inserting the analyzer exit slit, by collecting images for different photoelectron energies. Depending on whether core level or valence band photoelectrons are used, one can perform microprobe photoelectron diffraction or ARPES measurements. 

The angular acceptance of PEEM, determined in major part by the objective lens, depends on the electron kinetic energy. The full range of maximum allowed angles corresponds to about 11~\AA$^{-1}$ in reciprocal space units. Such an angular acceptance covers the first Brillioun zone of most materials.

\subsection{Applications}

Exfoliated graphene, the most frequently used type of graphene in prototypical devices, 
has attracted intense scientific interest concerning the relationship between morphology and transport properties. Since single layer crystals extend over lengths of at most few tens of microns, the study of their properties requires a microscopy approach. 

The SPELEEM measurements on exfoliated graphene flakes faced numerous challenges. First, the SiO$_2$ support, needed to locate the flakes with an optical microscope, necessitates to deposit grounding electrodes to avoid charging effects during experiments. A second and more severe complication is due to the corrugations in the graphene layer, which conforms to the rough morphology of the SiO$_2$ substrate. Such corrugations impose a serious obstacle to ARPES measurements, since they destroy photoelectron coherence. The evident short-range roughness of SiO$_2$-supported graphene is manifested by a pronounced broadening of all ARPES features~\cite{knox08}. Nonetheless, momentum distribution curve analysis demonstrates a massless fermionic dispersion of the $\pi$ band close to the Fermi level, with a Fermi velocity approaching $1.0 \cdot 10^6$~m/s, comparable to the typical values found in the literature. 

The broadening due to corrugations is considerably reduced on suspended exfoliated graphene. The suspended  samples are prepared by etching small cylindrical cavities in the SiO$_2$ substrate, over which the films are not in contact with the substrate. Because of the absence of an interacting support, suspended graphene displays a smoother texture than supported graphene, resulting in a notable narrowing of the diffraction spots in LEED as well as a better resolved ARPES pattern~\cite{knox11}.
 
Broadening effects in ARPES can be circumvented when using a LEEM-PEEM microscope. This is done by carrying out independent $\mu$-LEED measurements of peak broadening versus increasing momentum transfer. Diffraction spot-profile analysis can be used to determine the scaling parameters describing roughness~\cite{LocaACSN10}, which is intimately related to the photoemission linewidth~\cite{knox11}. Notably, this approach enables to obtain the intrinsic ARPES linewidth and can be applied to determine the band structure of a variety of corrugated 2D systems. In the case of suspended graphene, by separating corrugation from lifetime broadening effects we confirmed that the electronic structure of suspended EG is that of ideal, undoped, graphene. Most importantly, these measurements validate the current picture that suspended graphene behaves as a marginal Fermi liquid, showing a quasiparticle lifetime that scales as $(E-E_F )^{-1}$, in accord with theoretical predictions for undoped graphene~\cite{DasSarmaPRB75}.

\section{Angle resolved PEEM imaging}

\subsection{Methodology of bright and darkfield imaging}

Normal XPEEM operation utilizes photoelectrons emitted along the surface normal by positioning the contrast aperture at the center of the ARPES pattern. This condition, which can be defined as the brightfield operation in analogy to LEEM, is illustrated in Fig.~\ref{fig:dfSketch}a. Conversely, darkfield operation, or df-PEEM, is implemented by selecting off-normal photoelectrons for image formation. Several approaches can be used to implement df-PEEM:

\begin{figure*}[t]
\begin{center}
\includegraphics[width=14.5cm]{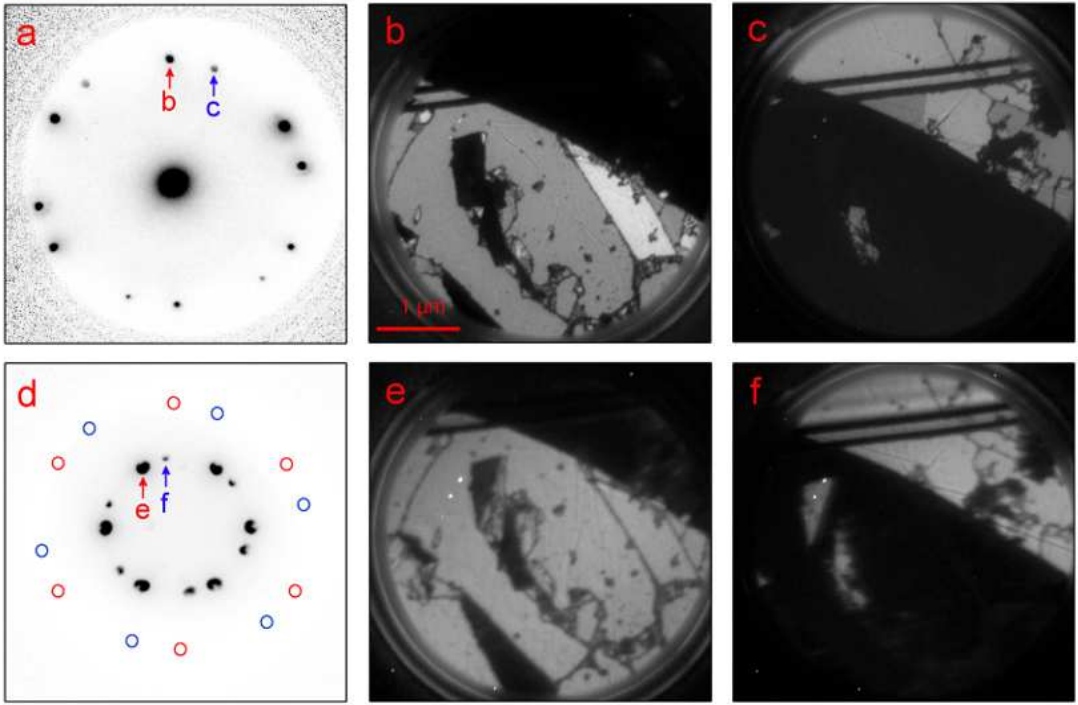}
\caption{a) $\mu$-LEED pattern of the HOPG surface from a region with two domains of different azimuthal orientation.
b) and c) df-LEEM images acquired using the corresponding 1st order LEED spots at 45~eV electron energy.
d) ARPES pattern from the same region at electron kinetic energy 76.0~eV and photon energy 80.0 eV.
At this photon energy, we access the $\Gamma$-K-M-$\Gamma$ plane of graphite. The blue and red circles
mark the positions of the LEED spots for the two domains as reciprocal space reference.
e) and f) df-XPEEM images acquired at the K-point of the corresponding rotational domains. 
Acquisition time is about 4~minutes per image.}
\label{fig:HOPG}
\end{center}
\end{figure*}

\begin{itemize}
\item To position the contrast aperture onto the desired k-point in reciprocal space, as shown in Fig.~\ref{fig:dfSketch}b. This method has the disadvantage of selecting beams that are displaced from the microscope optical axis. For large displacements, spherical aberrations may become relevant, resulting in poor image quality. A further disadvantage, specific to this instrument, is given by the imprecise mechanical control of the contrast aperture.

\item To deflect the beam after the objective lens, if possible\footnote{In the case of our instrument we used the image equalizers in the beam splitter of our microscope.}, in order to let the diffraction feature of interest pass through the aperture, which is kept fixed. This method, which is illustrated in Fig.~\ref{fig:dfSketch}c, is far more practical than the others, since the beam deflectors can be remotely controlled and their settings can be precisely calibrated in reciprocal space units. However, the photoelectrons that are far away from the objective lens optical axis induce aberrations. Thus, application of this method shall be limited to the case of a relatively small reciprocal space window centered around normal emission.

\item To modify the sample tilt in order to let the diffraction feature of interest pass through the contrast aperture, which is kept fixed. Although this approach is rather cumbersome, it provides an important advantage. As shown in Fig.~\ref{fig:dfSketch}d, the selected photoelectrons travel very close to the objective lens optical axis. Further, since the selected diffraction beam continues to propagate along the imaging column optical axis, there is no need for further alignment. Aberrations are minimized, leading to optimal image quality. The microscope performance in such an operation will be shortly illustrated in a case study on the orientational domains of the HOPG surface. This method is in principle equivalent to that of tilting the illumination beam in darkfield LEEM.

\end{itemize}

In practical cases, a combination of the second and third approach is most appropriate, especially when operating far from normal emission conditions.


\subsection{Lateral resolution}

In order to demonstrate the df-XPEEM operation we have chosen a HOPG specimen. 
The HOPG surface presents micron-sized domains, all with the (0001) orientation, 
but showing different azimuthal alignments. The electronic structure is nearly identical to that of graphite, with an almost linear dispersion of the  $\pi$ band at the Dirac points close to the Fermi level. The simple and well-known electronic band structure of graphite, showing no states at the $\Gamma$ point and sufficiently high density at the Dirac cones is ideally suited for the application of the darkfield method and provides a clear model for performance.

A $\mu$-LEED pattern from a region with only two orientational domains is shown in Fig.~\ref{fig:HOPG}a. The angle between the two domains is measured to be $20^\circ \pm 1^\circ$. This results in a clear separation of the $\pi$ bands in proximity of the Fermi level. The distinct HK axes corresponding to the two domains are separated by about 0.58~\AA$^{-1}$, as shown by the ARPES pattern in Fig.~\ref{fig:HOPG}d. Therefore,  the $\pi$ bands are well-resolved within the angular cones defined by the two smaller contrast apertures of our instrument, spanning 0.51 and 0.34~\AA$^{-1}$ in reciprocal units.

In order to visualize the rotational domains, we first acquired df-LEEM images. 
Those that correspond to the first order LEED spots of their respective domains are shown in Figs.~\ref{fig:HOPG}b and \ref{fig:HOPG}c. Bright contrast in df-LEEM marks the regions contributing to the selected diffraction spot. Aside the notable imperfections on the crystal surface (scratches, faceted regions, etc), one can see regions of slightly higher or lower gray level, which suggest the presence of slight variations of the azimuthal alignment. The sharpest features measure about 20~nm, which sets an upper limit for the estimate of the lateral resolution.

\begin{figure*}[t]
\begin{center}
\includegraphics{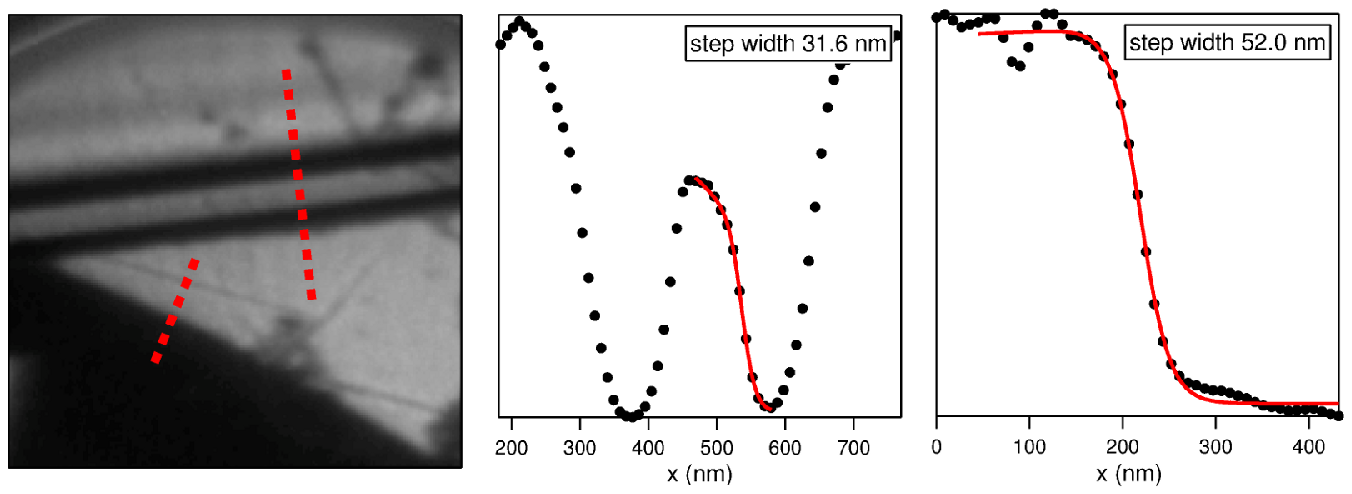}
\caption{On the left, a blow-up from Fig.~\ref{fig:HOPG} is seen. The intensity along the profiles marked with dashed lines
are shown in the two panels on the right. The sharpness of some features are indicated on the figure, as obtained using
a step-function fit.}
\label{fig:profile}
\end{center}
\end{figure*}

The respective df-XPEEM images of the same domains, obtained using the emission 
from the $\pi$ band in proximity to the $K$ point, are shown in Figs.~\ref{fig:HOPG}e and 
\ref{fig:HOPG}f. For each image, the photon energy was 80~eV, and the electron kinetic energy was 76~eV, corresponding to a binding energy of 0.9~eV. The medium contrast aperture with 0.51~\AA$^{-1}$ angular acceptance was here utilized in order to optimize the compromise between image statistics and lateral resolution. The darkfield images shown in the figure were obtained by averaging 16 drift corrected images, each acquired with an exposure time of 15 s. In order to get the desired features of the $\pi$ band through the diffraction aperture, we mechanically tilted the sample by about $1.8^\circ$, evaluated by assuming that the pivot point of the manipulator is on the sample plane. Final adjustments were accomplished by using the electron beam deflectors located in the beam splitter of our microscope.

\begin{figure*}[t]
\begin{center}
\includegraphics[width=15cm]{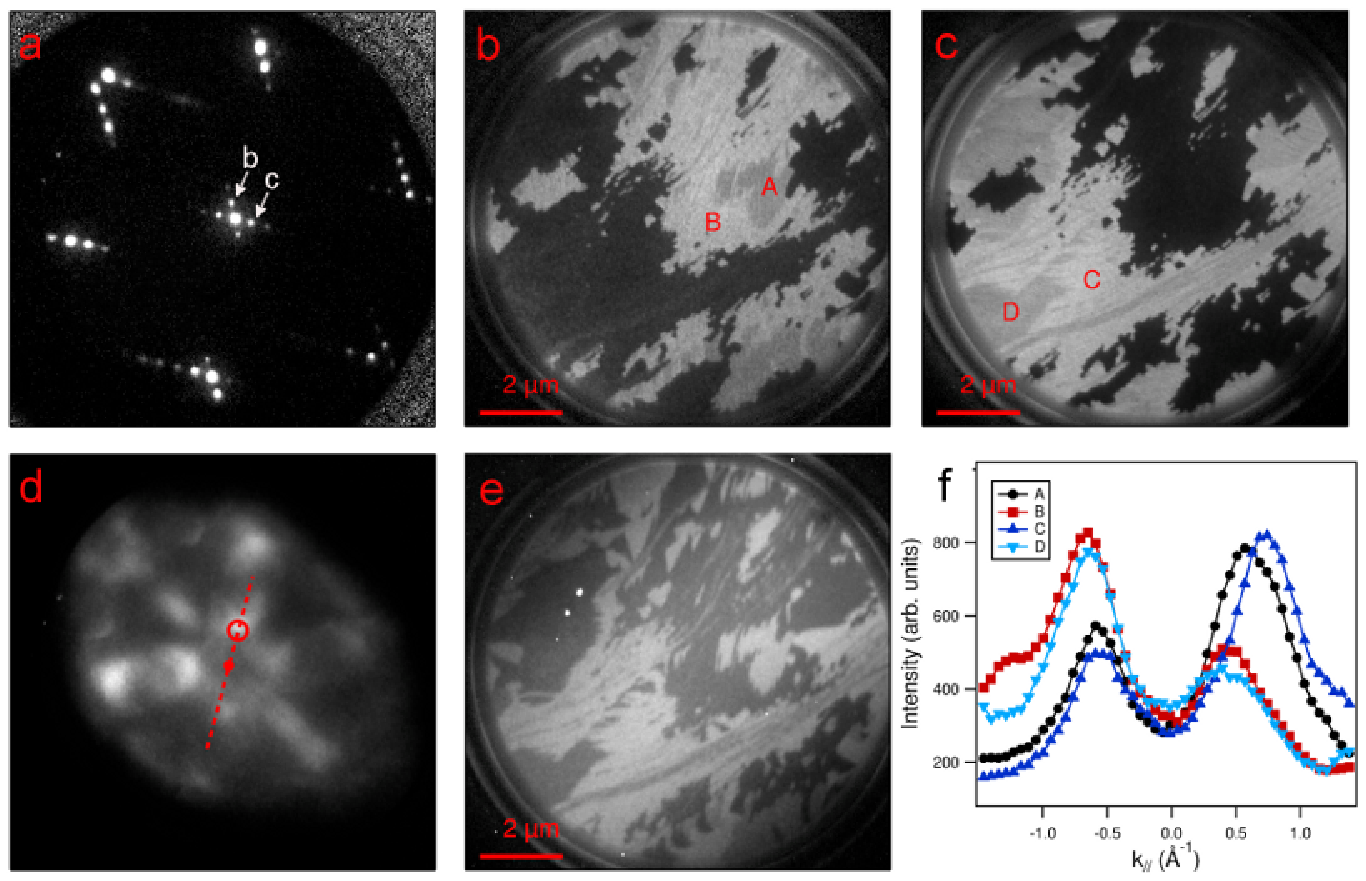}
\caption{($1\times1\times12$)-O/W(110): a) LEED pattern showing both rotational domains. 
b), c) df-LEEM images acquired using the superstructure spots marked on the LEED pattern. 
The labels A, B, C, D denote the regions used in the plot in the last panel. d) ARPES pattern from 
a single domain (A) at photon energy 38~eV, and electron kinetic energy 33.2~eV 
corresponding to 1.7~eV binding energy. e) df-XPEEM on the k-point marked 
by the  empty circle on the ARPES pattern. f) Intensity of each domain measured in df- 
XPEEM mode by scanning the $k_{\parallel}$ along the dashed line in panel (d).}
\label{fig:OW110}
\end{center}
\end{figure*}

As a general outline, the df-LEEM (Figs.~\ref{fig:HOPG}b and \ref{fig:HOPG}c) and df-XPEEM (Figs.~\ref{fig:HOPG}e and 
\ref{fig:HOPG}f) images show the presence of the same two orientational domains. However, a closer
inspection reveals subtle differences, which can be traced back to the slight variations of the grain orientations. Importantly, the angle-resolved photoemission intensity around the Dirac point covers a larger area in reciprocal space as opposed to the sharp LEED spots. Therefore, contrast due to small variations in azimuthal angle is smeared out in the df-PEEM images. Furthermore, one can also notice a small grain in the df-PEEM images, which gives no intensity in the df-LEEM images. This is also not surprising for a domain tilted out of the contrast aperture, but nevertheless gives intensity in df-PEEM due to the broad k-space features in ARPES.

It is important to note the small angle required to get to the edge of the first Brillouin zone. The actual photoelectron emission angle is 21.5$^\circ$, which is readily identified from the $\Gamma-K$ distance, 1.703~\AA$^{-1}$, and the perpendicular momentum of a photoelectron with kinetic energy of 76 eV. However, this angle shrinks greatly after the 18~kV accelerating field of the objective lens. By taking this into account, we find the required tilt angle to be 1.45$^\circ$, consistent with the applied mechanical tilt. The small difference is likely due to the uncertainty in locating the pivot point of the mechanical tilt in the sample manipulator, due to imperfections in the manipulator mechanics.

The lateral resolution was estimated from the width of the sharpest edges in the image, as has been illustrated in Fig.~\ref{fig:profile}. Profiles across selected edges are displayed in the same figure with the corresponding step-function fits. Averaging over several edges in order to be less sensitive to variations in the boundary morphology, we obtain a value of $40 \pm 5$~nm. This is an excellent figure considering that the best lateral resolution measured in XPEEM operation is about 30~nm~\cite{locatelli11}.

\subsection{Applications: (1$\times$12)-O/W(110)}

We have shown in the previous section how the df-XPEEM method can be applied to image different crystal grains, such as azimuthal domains in HOPG. A similar approach is obviously not limited to polycrystalline specimens, but can be extremely useful to image domains in a wide variety of superlattices. Furthermore, by collecting photoelectrons at selected emission angles, df-XPEEM can provide other useful information related to the local order near the emitter or the local surface electronic structure. In the following, we will provide the example of a well studied model system for surface science, oxygen on W(110).

Oxygen adsorption on W(110) has been under study since long~\cite{engel75, bauer78}. In the high coverage regime, a ($1\times12$) superstructure is observed in LEED with an oxygen coverage of about 1.08~ML~\cite{bauer78}. The adsorption site was later identified to be near the triply-coordinated positions, and the superstructure was attributed to alternating site-exchange domains in STM measurements~\cite{johnson93}.
The oxygen adsorption site determination was confirmed also in photoelectron diffraction measurements \cite{ynzunza99, ynzunza00}. Notably, all studies to date have considered structures with two (rotational) domains allowed.  Indeed, the triply-coordinated adsorption site, obtained by displacing the oxygen atoms from the hollow site towards the threefold site along $[1\bar{1}0]$, should give rise to two domains due the broken mirror symmetry along this direction.

In order to grow the high coverage oxygen layer, the clean W(110) surface was exposed to $1\times10^{-6}$~mbar of molecular oxygen at 1100~C$^{\circ}$ for 15 minutes, which resulted in the formation of a sharp ($1\times12$) LEED pattern. As shown in Fig.~\ref{fig:OW110}a, the extra spots produced by the two rotational domains of the oxygen phase are clearly visible around the central (00) peak. The df-LEEM images corresponding to the two superstructure spots marked in the LEED pattern are shown in Figs.~\ref{fig:OW110}b and~\ref{fig:OW110}c. Careful inspection reveals that each of such rotational domains breaks into two sub-domains, which appear with slightly different brightness in the df-LEEM images (see for example the regions labeled A and B in the Fig~\ref{fig:OW110}b).

The ARPES pattern from one of such domain is shown in Fig.~\ref{fig:OW110}d. The image was obtained using 38~eV photons, at electron kinetic energy of 33.2 eV. The broad feature identified by the red circle is intimately related to the oxygen valence band and could not be observed on the clean W(110) surface. The same state appears in normal emission at about 0.5~eV binding energy~\cite{feydt99}, and disperses along the $\bar{\Gamma}-\bar{S}$ direction (dashed line in Fig.~\ref{fig:OW110}d) for increasing binding energy.  At around 1.7 eV binding energy and 0.58~\AA$^{-1}$ away from normal emission, the density of states is sufficiently intense to allow XPEEM imaging. The corresponding df-XPEEM image is shown in Fig.~\ref{fig:OW110}e, obtained by averaging 30 drift-corrected images each acquired with a 10~s exposure time. Surprisingly, the df-XPEEM image shows different domains from both df-LEEM images. The new domains are formed by the different rotational sub-domains A+B and B+D, thus showing that the rotational domains are intermixed in the darkfield XPEEM image 

The experimental data can be rationalized on the basis of the 4 different sub-domains identified by darkfield LEEM and XPEEM data. We propose that the simplest adsorption model for the oxygen high coverage phase shows both mirror symmetries broken, thus leading to 4 structural domains. This condition is achieved when the oxygen adsorption site is slightly shifted away from the triply-coordinated site along $[001]$. Thus, the remaining mirror symmetry of the system is broken, introducing two additional domains that are distinguished by the sign of this shift. Note that none of the photoelectron diffraction studies had considered such a displacement of the oxygen atoms~\cite{ynzunza99, ynzunza00}. However, a recent DFT calculation on the ($1\times2$) oxygen on W(110) has found a shift of about 0.05~\AA\ in the $[001]$ direction~\cite{mentes08}. Assuming that the situation is qualitatively similar also for the high oxygen coverage case, the small shift explains the small contrast in df-LEEM images between domains A and B, or between C and D, as seen in Figs.~\ref{fig:OW110}b,c. Most importantly, we conclude that df-XPEEM allows to image the different adsorption site domains, thus enabling to achieve sensitivity to the symmetry of the adsorption site.

\subsection{Local spectroscopy in df-PEEM}
In order to map the local electronic structure, one can think of measuring the photoemission intensity inside a given region of interest in real-space darkfield XPEEM images acquired while scanning the photoelectron kinetic energy or the photoelectron parallel momentum. In the latter case, this is accomplished most reproducibly and conveniently by deflecting of the photoelectron beam before the contrast aperture.

The demonstration of this method is given for the case of O/W(110). In order to select the desired ARPES feature for imaging, the sample tilt was kept fixed, and the particular reciprocal space point was deflected into the contrast aperture by using the beam deflectors. In a similar way, spatially-resolved scans of the reciprocal space were obtained by varying the deflection angle and acquiring images at each point. Fig.~\ref{fig:OW110}f shows a plot of darkfield ARPES intensity in each of the four oxygen domains along a reciprocal-space profile marked by the dashed line  in Fig.~\ref{fig:OW110}d. The direction of the profile line was chosen to coincide with the direction of dispersion of the particular band feature. The exposure time for each image within the scan was 10~s. It is evident that the darkfield XPEEM method has a drastically different sensitivity to the broken symmetries in structural domains, which can be increased by choosing the proper band feature.

\section{Perspectives and conclusions}

XPEEM is a versatile technique for imaging surfaces and has its strengths in the large number of different and complementary methods that it provides. Here, we have illustrated the ARPES capabilities of PEEM, and the usefulness of their application in a variety of microscopy settings. On the one hand, the modest energy resolution in XPEEM puts it at a disadvantage compared to laterally-averaging high energy resolution devoted ARPES setups. On the other hand, its very high lateral resolution makes XPEEM a unique ARPES facility. As demonstrated by this study, angle resolved PEEM imaging can be done in normal emission, but also off normal (df-XPEEM). This ability is of crucial relevance for a number of future applications in material science. 

The potential of df-XPEEM is well illustrated by the imaging of structural domains in HOPG and on O/W(110). In the latter, df-PEEM was shown to provide complementary information with respect to the closely related df-LEEM method, enabling us to gain sensitivity to the adsorption site symmetry. The measured lateral resolution in the darkfield XPEEM operation is about 40~nm, comparable to that in normal emission XPEEM, and slightly lower than in df-LEEM operation as expected. 
The angular resolution, given by the contrast aperture, is 0.34~\AA$^{-1}$.
The combination of df-LEEM with df-PEEM creates a powerful tool that can sort out the structural and 
electronic heterogeneities on a surface with very high spatial resolution.

Beyond identifying the electronic band features corresponding to structurally inequivalent regions, 
the importance of angle-resolved XPEEM with high lateral resolution can be appreciated also in cases involving electronic heterogeneities in structurally homogeneous systems. In this category, a recent example was the theoretical account~\cite{hwang07} and the experimental observation~\cite{martin07} of spontaneously formed electron and hole puddles in exfoliated graphene.
The existence of these naturally-doped regions, with dimensions of about 100~nm, has profound implications on the transport properties of graphene. Clearly, darkfield XPEEM imaging near the Dirac point, as applied to HOPG in the preceding pages, would allow high-resolution images sensitive to the position of the Dirac point with respect to the Fermi level, and hence to the
doping of the graphene sheet. Similar studies are in progress on epitaxial graphene on metal substrates~\cite{noteLocatelli}.

In conclusion, darkfield PEEM widens the capabilities of photoemission electron microscopy, further extending its sensitivity to the electronic structure. We envisage that darkfield XPEEM can be fruitfully applied when imaging inhomogeneous interfaces and polycrystalline surfaces \cite{dudin10}, in order to obtain high lateral resolution maps of the electronic band structure. Examples of potential applications range from role of electronic phase separation in complex oxides~\cite{dagotto01}, to electronic effects in graphene flakes~\cite{knox08, knox11}, and to Rashba effect at ferromagnetic surfaces~\cite{krupin05}.

\section{Acknowledgments}

We acknowledge E. Bauer for critical reading of the manuscript and several illuminating discussions on the subject.

\bibliographystyle{model1a-num-names}

\end{document}